\documentclass[preprint,superscriptaddress,amsmath,amssymb,prd,aps,showpacs,floatfix,nofootinbib]{revtex4-1}
\usepackage{graphicx}% Include figure filesa
\usepackage{dcolumn}% Aligntable columns on decimal point
\usepackage{bm}
\usepackage{mathrsfs}% bold math
\usepackage{hyperref}% add hypertext capabilities
\usepackage{amsmath}
\usepackage{amssymb}
\usepackage{bm}
\usepackage{color}
\usepackage{float}
\usepackage{dcolumn}
\usepackage{multirow}
\usepackage{changepage}
\hypersetup{pdftex,colorlinks=true,linkcolor=blue,citecolor=red,menucolor=black,urlcolor=blue,filecolor=blue}

\newcommand{\minv}{M_{\rm inv}}

\begin{document}
%\title{$\boldsymbol{\Xi_{bb}}$ and $\boldsymbol{\Omega_{bbb}}$ molecular states}
\title{Testing the origin of the ``$f_1(1420)$" with the $\bar K p \to \Lambda (\Sigma) K\bar K \pi$ reaction}
\date{\today}

\author{Wei-Hong~Liang}
\email{liangwh@gxnu.edu.cn}
\affiliation{Department of Physics, Guangxi Normal University, Guilin 541004, China}
\affiliation{Guangxi Key Laboratory of Nuclear Physics and Technology, Guangxi Normal University, Guilin 541004, China}

\author{E.~Oset}
\email{oset@ific.uv.es}
\affiliation{Department of Physics, Guangxi Normal University, Guilin 541004, China}
\affiliation{Departamento de F\'{\i}sica Te\'orica and IFIC, Centro Mixto Universidad de
Valencia-CSIC Institutos de Investigaci\'on de Paterna, Aptdo.22085,
46071 Valencia, Spain}
%             %  but any date may be explicitly specified

\begin{abstract}
  We study the $\bar K p \to Y K\bar K \pi$ reactions with $\bar K = \bar K^0, K^-$ and $Y=\Sigma^0, \Sigma^+, \Lambda$,
  in the region of $K\bar K \pi$ invariant masses of $1200-1550$ MeV.
  The strong coupling of the $f_1(1285)$ resonance to $K^* \bar K$ makes the mechanism based on $K^*$ exchange very efficient
  to produce this resonance observed in the $K\bar K \pi$ invariant mass distribution.
  In addition, in all the reactions one observes an associated peak at $1420$ MeV which comes from the $K^* \bar K$ decay mode of the $f_1(1285)$
  when the $K^*$ is placed off shell at higher invariant masses.
  We claim this to be the reason for the peak of the $K^* \bar K$ distribution seen in the experiments which has been associated to the ``$f_1(1420)$" resonance.
\end{abstract}

%\pacs{11.30.Er, 12.39.-x, 13.25.Hw}% PACS, the Physics and Astronomy
\maketitle

%\tableofcontents

\section{Introduction}
\label{sec:intro}
Kaon beams are called to play a relevant role in studies of hadron dynamics.
So far kaon beams in the GeV range are available at the Japan Proton Accelerator Research Complex (J-PARC) \cite{XieLiang1, XieLiang2}.
Kaon beams of low energy from $\phi$ decay are available in the Phi-Factory at Frascati \cite{XieLiang3, XieLiang4}.
The COMPASS collaboration at CERN can also select events coming from kaon induced reactions \cite{Compass,Comexo,Wallner}.
In addition the planned $K_L$ factory at Jefferson Lab has passed the first steps for approval \cite{Jeff, proposal}.
The main aim of this latter factory is the study of strange hadron spectroscopy,
but its potential for non-strange hadron studies is also relevant.
In this sense, having this facility in mind, a proposal was made in Ref.~\cite{XieLiang}
to produce the $f_0(980)$ and $a_0(980)$ resonances in $\bar K$ induced reactions on protons,
which should shed light on the much debated nature of these resonances.

In the present work we make a proposal that shows the usefulness of such facilities to provide relevant information on the nature of the ``$f_1(1420)$" resonance,
which is catalogued in the PDG \cite{pdg} as a standard resonance.
This resonance has been observed in many very high energy reactions,
as proton or pion induced, $e^+e^-$, $\gamma \gamma$ and $J/\psi$ decays,
and exceptionally in one experiment with a $K^-$ induced reaction at $32.5 ~{\rm GeV}/c$ in Serpukhov \cite{Serpu}.
One of the peculiar properties of this resonance is that it is only seen in the $K\bar K\pi$ decay mode, supposed to be $K^* \bar K +c.c.$,
with a small fraction seen in the $a_0(980) \pi$ channel.
However, the right to be catalogued as a standard resonance was challenged in Ref.~\cite{ViniFran},
where it was proved that in such reactions the $f_1(1420)$ peak appeared naturally as the $K^* \bar K +c.c.$ decay mode of the $f_1$(1285) resonance,
which showed two peaks in the $K\bar K \pi$ decay channel,
one at the nominal $f_1(1285)$ mass and another one around $1420$ MeV, where the $K^* \bar K$ channel with $K^*$ on shell becomes open.
The small fraction of $a_0(980)\pi$ decay was also shown in Ref.~\cite{ViniFran} to correspond to a triangle singularity decay mode of the $f_1(1285)$.

In the present work we show that the $f_1(1420)$ peak can also be produced in the present and future lower energy Kaon Facilities with reactions that one has under control theoretically,
such that their experimental search and comparison with the theoretical predictions can shed much valuable light on the origin and nature of the ``$f_1(1420)$" peak.
For this purpose we propose several reactions,
all tied among themselves,
$K^- p \to \Lambda (\Sigma^0) K^0 K^- \pi^+$, and $\bar K^0 p \to \Sigma^+ K^0 K^- \pi^+$, for which we make evaluations of the cross section,
with the assumption that the $f_1(1420)$ is only a manifestation of the $K^* \bar K + c.c.$ decay mode of the $f_1(1285)$.
We make absolute predictions of the cross section for the three reactions,
and the strength of this cross section around the $f_1(1285)$ and $f_1(1420)$ peaks results
as a consequence of the assumed nature of the $f_1(1285)$
as a dynamically generated resonance of the pseudoscalar-vector meson interaction \cite{Lutz,Luis,Geng,Juan}.
The measurement of the cross section can also provide support for this latter hypothesis,
but the peak of the $f_1(1420)$ is simply a consequence of the coupling of the $f_1(1285)$ to the $K^* \bar K$ channel,
which is corroborated explicitly by the experimental $\bar K K\pi$ decay mode of the $f_1(1285)$ (see also a theoretical description of this decay mode in Ref.~\cite{Aceti}).

\section{Formalism}
\label{sec:Formalism}
The $f_1(1285)$ axial vector meson has the quantum numbers $I^G (J^{PC})=0^+ (1^{++})$.
Using chiral Lagrangians for pseudoscalar-vector interaction \cite{Birse} and a chiral unitary approach in coupled channels,
the axial vector mesons emerge as a consequence of the interaction \cite{Lutz,Luis,Geng,Juan}.
In particular the $f_1(1285)$ is the cleanest example, appearing in a single channel $K^* \bar K - c.c.$.
Concretely the state is given, with the isospin convention $(K^+, K^0)$, $(\bar K^0, K^0)$ and $C (K^{*+})= -K^{*-}$, etc., by
\begin{equation}\label{eq:f1}
  | f_1(1285) \rangle = - \frac{1}{2} \, | K^{*+} K^- + K^{*0} \bar K^0 - K^{*-} K^+ - \bar K^{*0} K^0 \rangle.
\end{equation}
The coupling of the state to this $K^* \bar K -c.c.$ combination in $s$-wave obtained from the residues of the $t_{f_1, f_1}$ amplitude at the pole
\begin{equation}\label{eq:coup}
  t_{f_1, f_1}= \frac{g^2_{f_1, K^*\bar K}}{z-z_R},
\end{equation}
with $z$ the complex energy, and $z_R$ the complex pole position, is given by
\begin{equation}\label{eq:gf1}
  g_{f_1, K^* \bar K }=7219 ~ {\rm MeV} ~[{\color{red} 16}]; ~~~~
  g_{f_1, K^* \bar K} =7230 ~ {\rm MeV} ~[{\color{red} 15}].
\end{equation}
%\begin{equation}\label{eq:gf1}
%  g_{f_1, K^* \bar K }=7219 ~ {\rm MeV} ~\cite{Geng}; ~~~~
%  g_{f_1, K^* \bar K} =7230 ~ {\rm MeV} ~\cite{Luis}.
%\end{equation}
%
We shall take the second value in our calculations.
The $t_{f_1, K^{*+} K^-}$ amplitude is then given by
\begin{equation}\label{eq:amf1}
  -\frac{1}{2} \, g_{f_1, K^* \bar K} \ \vec \epsilon \;(f_1) \cdot \vec \epsilon \;(K^*),
\end{equation}
with $\vec \epsilon$ the polarization vector, similarly with the other components.

\begin{figure}[tb]
\begin{center}
\includegraphics[scale=0.48]{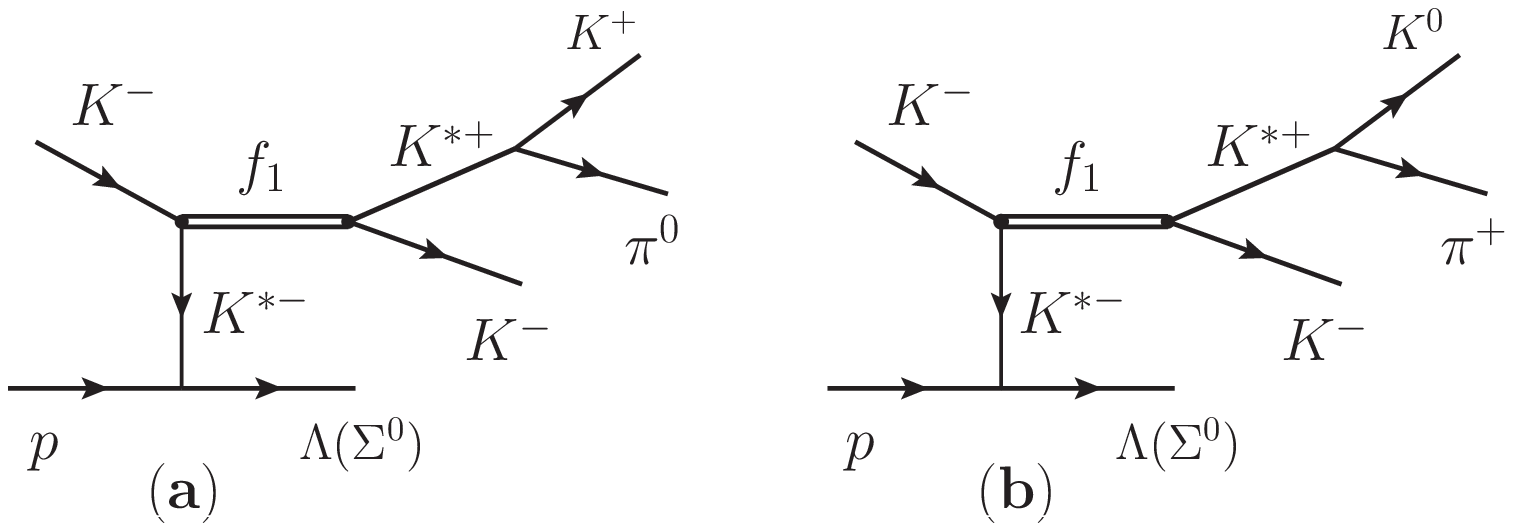}
\includegraphics[scale=0.48]{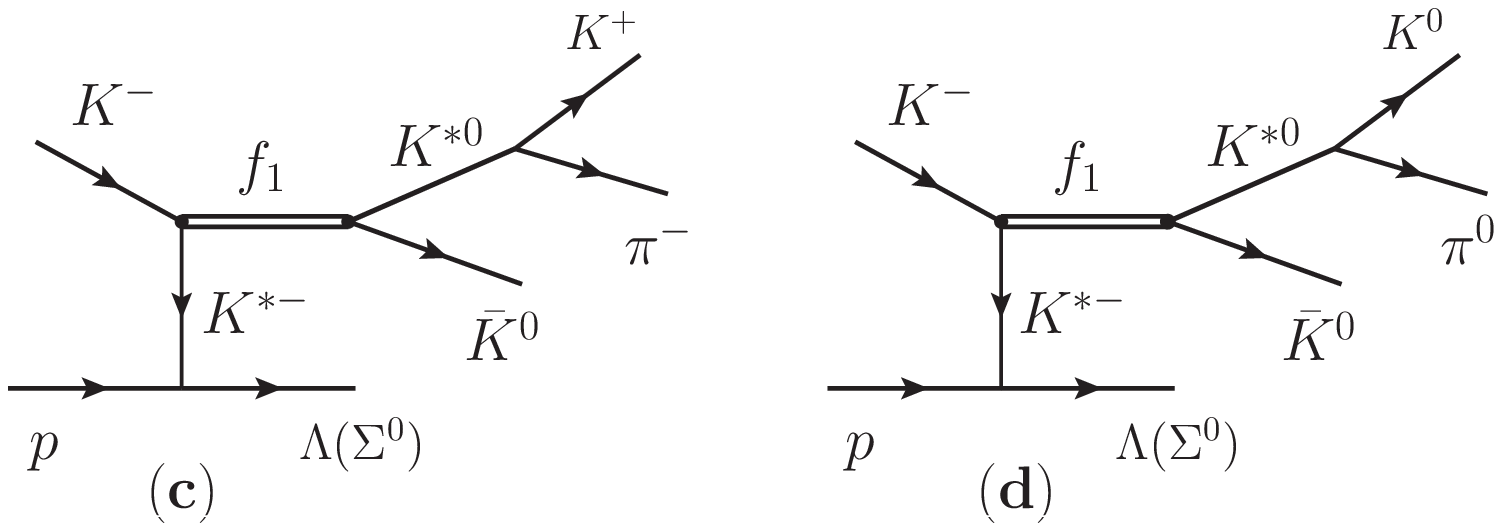}\\
\vspace{0.2cm}
\includegraphics[scale=0.48]{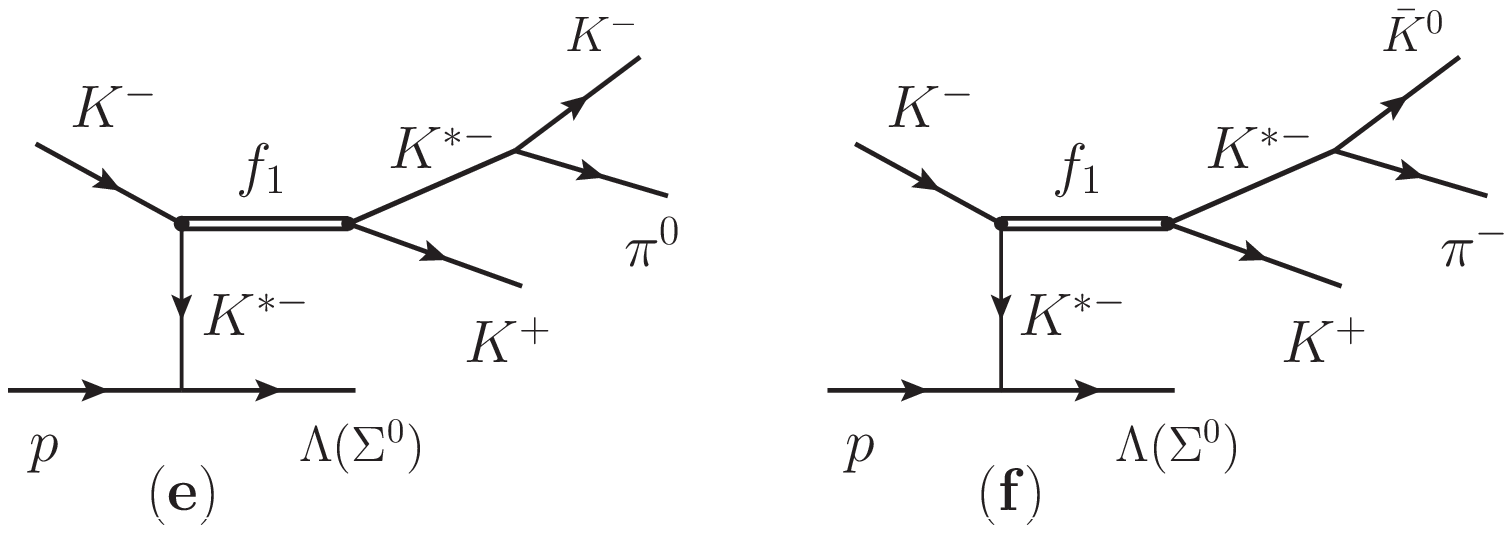}
\includegraphics[scale=0.48]{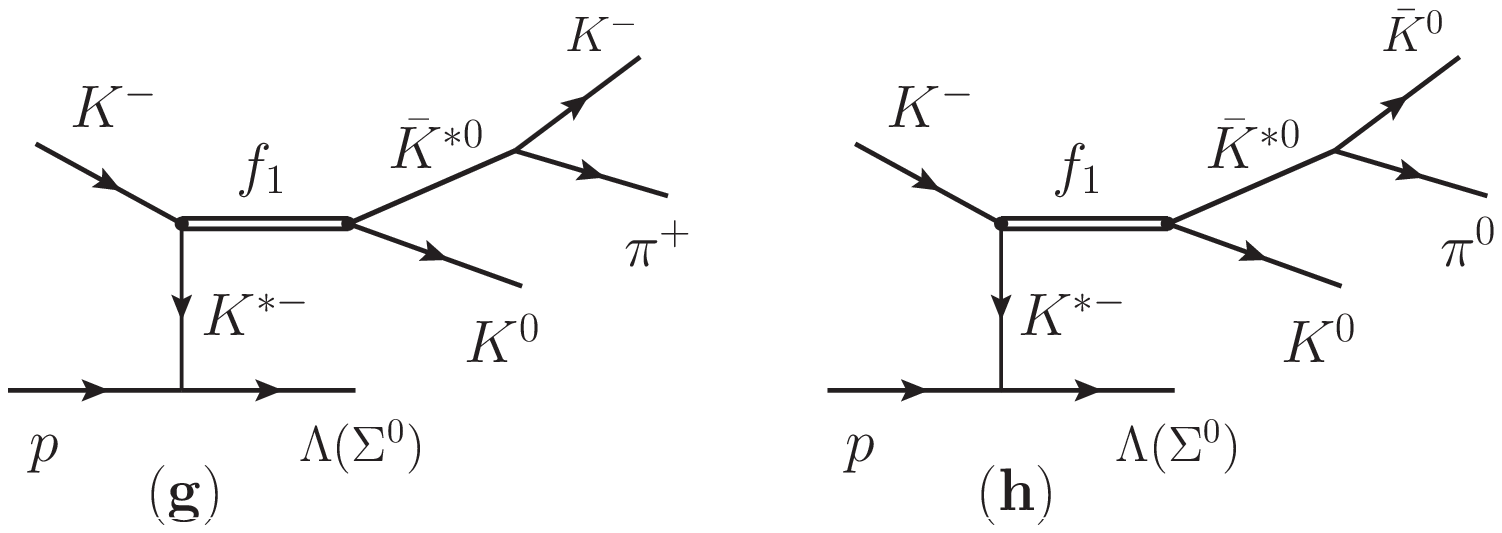}
\end{center}
\vspace{-0.4cm}
\caption{The mechanism for $K^- p \to \Lambda (\Sigma^0) f_1  \to \Lambda (\Sigma^0) K \pi \bar K$ reactions.}
\label{Fig:1}
\end{figure}

In view of this, the $K^-p \to Y f_1  \to Y K\pi \bar K $ decay proceeds via the mechanism shown in Fig.~\ref{Fig:1}.
We should note that this mechanism is the same one used in Ref.~\cite{Xie} to study the $K^- p \to \Lambda f_1(1285)$ reaction,
but in that work the process stops in the $f_1$ production of Fig.~\ref{Fig:1} without looking into the specific decay channel of $K \pi \bar K$,
hence, the $f_1(1420)$ excitation was not addressed in that work.

We shall specify the $K^0 \pi^+ K^-$ decay channel of $f_1$,
and, hence, only the (b) and (g) diagrams of Fig.~\ref{Fig:1} contribute.

Apart from the $f_1$ coupling to $K^* \bar K$,
we need the coupling of $K^*$ to $K\pi$ which is given by the standard Lagrangian
\begin{equation}\label{eq:LVPP}
  \mathcal{L}_{VPP}= -ig\, \langle [P, \partial_\mu P]\, V^\mu \rangle,
\end{equation}
with $P, V$ the standard SU(3) pseudoscalar and vector meson matrices
\begin{equation}
\label{Pmatrix}
P =
\left(
\begin{array}{ccc}
 \frac{1}{\sqrt{2}} \pi^0 + \frac{1}{\sqrt{6}} \eta & \pi^+ & K^+  \\[0.1cm]
\pi^- &  - \frac{1}{\sqrt{2}} \pi^0 + \frac{1}{\sqrt{6}} \eta & K^0\\[0.1cm]
K^- & \bar{K}^0 & -\frac{2}{\sqrt{6}} \eta  \\
\end{array}
\right),
\end{equation}
\begin{equation}
\label{Vmatrix}
V =
\left(
\begin{array}{ccc}
\frac{1}{\sqrt{2}} \rho^0 + \frac{1}{\sqrt{2}} \omega & \rho^+ & K^{* +}  \\[0.1cm]
\rho^- & -\frac{1}{\sqrt{2}} \rho^0 + \frac{1}{\sqrt{2}} \omega & K^{* 0}  \\[0.1cm]
K^{* -} & \bar{K}^{* 0} & \phi  \\
\end{array}
\right),
\end{equation}
with $g=M_V/2f_\pi$ ($M_V=800$~MeV, $f_\pi=93~$ MeV).

On the other hand, we need the $K^* BB$ vertex.
In chiral dynamics this vertex is given in terms of the Lagrangian \cite{Kaiser,RamosVec} \footnote{correcting a misprint in Ref.~\cite{Kaiser}.}
\begin{eqnarray}
\label{eq:Lvbb}
\mathcal{L}_{BBV} = g\, \Big( \langle \bar{B} \gamma_{\mu} [V^{\mu},B] \,\rangle +
\langle \bar{B} \gamma_{\mu} B \rangle \langle V^{\mu} \rangle \Big) \, ,
\end{eqnarray}
with the SU(3) baryon matrix $B$ given by
\begin{equation}
\label{eq:Bmatrix}
B =
\left(
\begin{array}{ccc}
\frac{1}{\sqrt{2}} \Sigma^0 + \frac{1}{\sqrt{6}} \Lambda & \Sigma^+ & p \\[0.1cm]
\Sigma^- & - \frac{1}{\sqrt{2}} \Sigma^0 + \frac{1}{\sqrt{6}} \Lambda & n \\[0.1cm]
\Xi^- & \Xi^0 & - \frac{2}{\sqrt{6}} \Lambda
\end{array}
\right).
\end{equation}

Since we will work with relatively low $\bar K$ energies,
we will take the spatial components of the $K^*$, $\epsilon^i$,
neglecting the $\epsilon^0$ component, which has been shown to be a very good approximation even for relatively large $K^*$ momenta \cite{SakaiRamos}.
The relevant couplings are then given by
\begin{eqnarray}\label{eq:vertex}
  V_{\bar K^{*0}, K^- \pi^+} &=& g \; \vec\epsilon \,(\bar K^{*0}) \cdot (\vec p_{K^-}- \vec p_{\pi^+}), \nonumber\\
  V_{K^{*+}, K^0 \pi^+} &=& -g \; \vec\epsilon\,(K^{*+}) \cdot (\vec p_{K^0}- \vec p_{\pi^+}), \\
  V_{BBV} &=& g\; \langle \bar u | \gamma^\mu  | u  \rangle \; C_B, \nonumber
\end{eqnarray}
with
\begin{equation}\label{eq:cB}
C_B=\left\{
             \begin{array}{cl}
             \dfrac{1}{\sqrt{2}}, & {\rm for}~ pK^{*-} \to \Sigma^0; \\[0.1cm]
             \dfrac{1}{\sqrt{6}}, & {\rm for}~ pK^{*-} \to \Lambda;\\[0.1cm]
             1, & {\rm for}~ pK^{*0} \to \Sigma^+.
             \end{array}
\right.
\end{equation}
A different $BBV$ vertex is used in Ref.~\cite{Xie} following Refs.~\cite{Doring,Ronchen}, given by
\begin{equation}\label{eq:VbbvXie}
  V'_{BBV}= g\, (1+2\alpha)\,\langle \bar u | [\gamma^\mu + i \dfrac{\kappa}{M+M'}\, \sigma^{\mu\nu}\, (p'-p)_\nu] | u \rangle,
\end{equation}
where $M, M', p, p'$ are the masses and momenta of the incoming and outgoing baryons,
with $\alpha =1.15$ and $\kappa=2.77$ according to Refs.~\cite{Holinde,XieZou} and a slightly different value of $g$.
The magnetic $\sigma^{\mu\nu}$ term of Eq.~\eqref{eq:VbbvXie} is usually neglected in chiral dynamics calculations since the momentum transfer is small,
but in this case it is not negligible and we shall take it into account.
The coupling of Eq.~\eqref{eq:VbbvXie} seems much bigger than that of Eqs.~\eqref{eq:vertex}, \eqref{eq:cB}
but they are accompanied with form factors which are normalized differently.
In the formalism of Eq.~\eqref{eq:VbbvXie} the form factor is normalized to unity when $q^2=(p'-p)^2 =m^2_{K^*}$,
while in Eqs.~\eqref{eq:vertex}, \eqref{eq:cB} it is normalized to unity for $q^2=0$.
We shall take a form factor of the type
\begin{equation}\label{eq:FF}
  \dfrac{\Lambda^2}{\Lambda^2-q^2}~~{\rm for~Eqs.~\eqref{eq:vertex}, \eqref{eq:cB}};~~~~ \dfrac{\Lambda^2-m^2_{K^*}}{\Lambda^2-q^2}~~{\rm for~Eq.~\eqref{eq:VbbvXie}},
\end{equation}
and we shall see the difference between the results with the two approaches, which we will accept as uncertainties of our results.

With the ingredients discussed above, the amplitude for the diagram of Fig.~\ref{Fig:1}(b) is given by
\begin{eqnarray}\label{eq:Amp}
  \tilde{t} &=& - \dfrac{1}{4}\, g^2\, g^2_{f_1, K^* \bar K} \, C_B \;
  \langle \bar u | [\gamma^i + i \dfrac{\kappa}{M+M'}\, \sigma^{i\nu}\, (p'-p)_\nu] | u \rangle \, (p_{K^0}-p_{\pi^+})^i  \nonumber\\[0.1cm]
   && \times \dfrac{1}{\minv^2 (K^0 \pi^+)-m^2_{K^*}+i\, m_{K^*} \,\Gamma_{K^*}}
       \; \dfrac{1}{\minv^2 (K^0 \pi^+ K^-)-M^2_{f_1}+i\, M_{f_1} \,\Gamma_{f_1}} \nonumber\\[0.1cm]
   && \times \dfrac{1}{q^2-m^2_{K^*} +i\, m_{K^*} \,\Gamma_{K^*}} \; \dfrac{\Lambda^2-m^2_{K^*}}{\Lambda^2-q^2}.
\end{eqnarray}
We should also sum coherently the contribution of the diagram of Fig.~\ref{Fig:1}(g).
Since $K^{*+}$ and $\bar K^{*0}$ are different particles,
in the limit of small $K^*$ width these diagrams do not interfere.
In addition, the different angles of $(\vec p_{K^0}- \vec p_{\pi^+})$ and $(\vec p_{K^-}- \vec p_{\pi^+})$ also make the interference smaller,
and then we neglect it.
As a consequence, and evaluating explicitly the $\langle \bar u | \gamma^i | u \rangle$, $\langle \bar u | \sigma^{i\nu} | u \rangle$ matrix elements
we obtain at the end,
\begin{eqnarray}\label{eq:Amp2}
 \overline{\sum} \sum |\tilde{t}|^2 &=&  \dfrac{1}{16}\, \left[g\, g(1+2\alpha)\, g^2_{f_1, K^* \bar K} \, C_B \right ]^2 \;
 \left(\dfrac{\Lambda^2-m^2_{K^*}}{\Lambda^2-q^2}\right)^2  \nonumber\\[0.2cm]
 && \times \left\{ \dfrac{1+2(1+\kappa)^2}{3} \, \left( \dfrac{\vec p^{\,2}}{4M^2} + \dfrac{\vec p^{\, \prime 2}}{4M^{\prime 2}} \right)
 -\dfrac{2}{3} \dfrac{|\vec p\,| \, |\vec p\,'|}{4MM'}\,[2(1+\kappa)^2 -1]\right\}  \nonumber\\[0.2cm]
 && \times \left| \dfrac{1}{\minv^2 (K^0 \pi^+ K^-)-M^2_{f_1}+i\, M_{f_1} \,\Gamma_{f_1}} \right|^2
    \cdot \left|\dfrac{1}{q^2-m^2_{K^*} +i\, m_{K^*} \,\Gamma_{K^*}} \right|^2  \nonumber\\[0.2cm]
   && \times \left\{ \dfrac{\lambda (\minv^2(K^0 \pi^+), m^2_{K^0}, m^2_{\pi^+})}{\minv^2 (K^0 \pi^+)}\,
   \left|\dfrac{1}{\minv^2 (K^0 \pi^+)-m^2_{K^*}+i\, m_{K^*} \,\Gamma_{K^*}}\right|^2  \right.  \nonumber\\[0.2cm]
   && + \left. \dfrac{\lambda (\minv^2(K^- \pi^+), m^2_{K^-}, m^2_{\pi^+})}{\minv^2 (K^- \pi^+)}\,
   \left|\dfrac{1}{\minv^2 (K^- \pi^+)-m^2_{K^*}+i\, m_{K^*} \,\Gamma_{K^*}}\right|^2 \right\},
\end{eqnarray}
where an angular average in $[(\vec p_K - \vec p_\pi) \cdot \vec p\,]^2$
and $[(\vec p_K - \vec p_\pi) \cdot \vec p\,']^2$ has been done,
and the cross section for the reaction is given by
\begin{eqnarray}\label{eq:crosec}
   &&\dfrac{{\rm d} \sigma}{{\rm d}\! \cos \!\theta \; {\rm d} \!\minv (K^0 \pi^+ K^-)} \nonumber \\[0.2cm]
   &=&  \dfrac{MM'}{s}\, \dfrac{p'}{p_{\bar K}} \, \dfrac{1}{(2\pi)^5} \, \dfrac{1}{32\, \minv (K^0 \pi^+ K^-)}
   \,\int {\rm d} \minv (K^- \pi^+)\, 2\,\minv (K^- \pi^+)\nonumber \\[0.2cm]
   && \times \int {\rm d} \minv (K^0 \pi^+)\, 2\,\minv (K^0 \pi^+)\; \overline{\sum} \sum |\tilde{t}|^2,
\end{eqnarray}
with the limits of the integration for $\minv (K^0 \pi^+)$ with fixed $\minv (K^- \pi^+)$ given in the PDG \cite{pdg},
and $p_{\bar K}$ the initial $\bar K$ momentum.

\begin{figure}[tb]
\begin{center}
\includegraphics[scale=0.6]{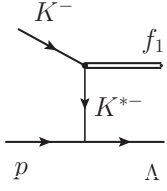}
%\hspace{0.9cm}
%\includegraphics[scale=0.7]{Fig1-b.eps}
\end{center}
\vspace{-0.7cm}
\caption{Mechanism for $f_1$ production in the $K^- p \to \Lambda f_1$ reaction.}
\label{Fig:2}
\end{figure}

We can tune the $\Lambda$ parameter to get an important datum
which is the integrated cross section for $f_1(1285)$ production in the $K^-p \to \Lambda f_1(1285)$
at $\sqrt{s}=3010$ MeV, $\sigma= 11\pm 3 \,\mu b$ \cite{Armenteros}.
The process can be described by the diagram of Fig.~\ref{Fig:2}.

Proceeding as before, we find now
\begin{equation}\label{eq:crosec2}
  \dfrac{{\rm d}\sigma}{{\rm d}\!\cos \!\theta}=\dfrac{M_p\, M_\Lambda}{8\pi}\, \dfrac{1}{s}\, \dfrac{p_{f_1}}{p_{\bar K}}\, \overline{\sum} \sum |t'|^2,
\end{equation}
where $p_{f_1}, p_{\bar K}$ are the $f_1$ and $\bar K$ momenta, and now ($C_B=\frac{1}{\sqrt{6}}$)
\begin{eqnarray}\label{eq:tprime}
 \overline{\sum} \sum |t'|^2 &=& \dfrac{1}{24}\, g^2_{f_1, K^* \bar K} \, g^2 (1+2\alpha) \, \left( \dfrac{\Lambda^2 - m^2_{K^*}}{\Lambda^2 -q^2}\right)^2
 \; \left|\dfrac{1}{q^2-m^2_{K^*} +i\, m_{K^*} \,\Gamma_{K^*}} \right|^2 \nonumber\\[0.2cm]
   &&\times \left\{ [1+2(1+\kappa)^2] \left( \dfrac{\vec p^{\,2}}{4M_p^2} + \dfrac{\vec p^{\, \prime 2}}{4M_\Lambda^{2}} \right)
 -\dfrac{2\,|\vec p\,| \, |\vec p\,'|}{4M_p M_\Lambda'}\,\cos \!\theta \,[2(1+\kappa)^2 -1]\right\}.
\end{eqnarray}

\section{Results}
\label{sec:results}
The first step is to use Eq.~\eqref{eq:crosec2} to obtain the cross section for $K^- p \to \Lambda f_1(1285)$.
By using Eq.~\eqref{eq:VbbvXie} for the $BBV$ coupling and a value of $\Lambda=1250$ MeV, we find $\sigma =11.07 \,\mu b$.
This result is compatible with those of Ref.~\cite{Xie} where a bigger $\Lambda$ is demanded
because extra powers of $\frac{\Lambda^2 - m^2_{K^*}}{\Lambda^2 -q^2}$ are used in the form factor.
If we use Eq.~\eqref{eq:crosec2} together with Eqs.~\eqref{eq:vertex}, \eqref{eq:cB} and the form factor of Eq.~\eqref{eq:FF} normalized to $1$ at $q^2=0$,
then we need a value $\Lambda=1900$ MeV and we get $\sigma=11.24 \,\mu b$.
We shall use these two versions of the $BBV$ vertex and evaluate the cross section for $K^0 \pi^+ K^-$ production of Eq.~\eqref{eq:crosec},
accepting the differences as uncertainties.
The results will be given with the coupling of Eq.~\eqref{eq:VbbvXie}.

\begin{figure}[tb]
\begin{center}
\includegraphics[scale=0.75]{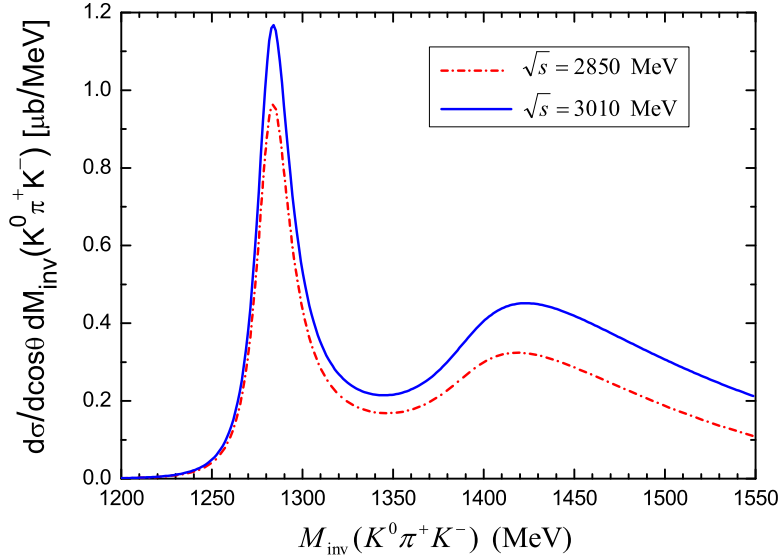}
\end{center}
\vspace{-0.7cm}
\caption{Differential cross section $\frac{{\rm d} \sigma}{{\rm d}\! \cos \!\theta \; {\rm d} \!\minv (K^0 \pi^+ K^-)}$ (with $\cos \!\theta =1$) for the $\bar K^0 p \to \Sigma^+ K^0 \pi^+ K^-$ reaction at $\bar K^0$ energy $\sqrt{s}=2850$ MeV or $3010$ MeV.}
\label{Fig:3}
\end{figure}

In Fig.~\ref{Fig:3} we show $\frac{{\rm d} \sigma}{{\rm d}\! \cos \!\theta \; {\rm d} \!\minv (K^0 \pi^+ K^-)}$ for $\cos \!\theta =1$
as a function of $\minv (K^0 \pi^+ K^-)$ for the $\bar K^0 p \to \Sigma^+ K^0 \pi^+ K^-$ at two $\bar K^0$ energies
corresponding to $\sqrt{s}=2850$ MeV and $\sqrt{s}=3010$ MeV.
We observe a clear peak at the $f_1(1285)$ nominal mass, corresponding to the $K\pi \bar K$ decay of the $f_1$ \cite{pdg},
but interestingly we find also a peak around $\minv (K^0 \pi^+ K^-)$ of $1420$ MeV.
This peak comes from the $f_1(1285)\to K^* \bar K$ when the $K^*$ becomes on shell.
We should see this peak as a combination of two factors: The increasing of $\minv (K^0 \pi^+ K^-)$ allows the intermediate $K^*$ to get on shell,
increasing the cross section,
but the tail of the $f_1(1285)$ tends to reduce the cross section with increasing $\minv$.
The result of it is the peak seen at $1420$ MeV which is hence the manifestation of the $K^* \bar K$ decay mode of the $f_1(1285)$ and did not come from any new resonance.

There is no need to show results for the $K^- p \to \Sigma^0 K^0 \pi^+ K^-$ reaction,
since neglecting the $\Sigma^0, \Sigma^+$ mass difference the cross section is just $\frac{1}{2}$ of the former one (see Eq.~\eqref{eq:cB}).
At this point, it is interesting to see what happens if we use the $BBV$ coupling of Eq.~\eqref{eq:cB}.
The results are very similar but about a factor of two smaller.
We accept this as uncertainty in our results.
Yet, the important thing is that the shape of the cross section is practically identical,
with about the same ratio of the strength at the $1420$ MeV and $1285$ MeV peaks.

\begin{figure}[tb]
\begin{center}
\includegraphics[scale=0.75]{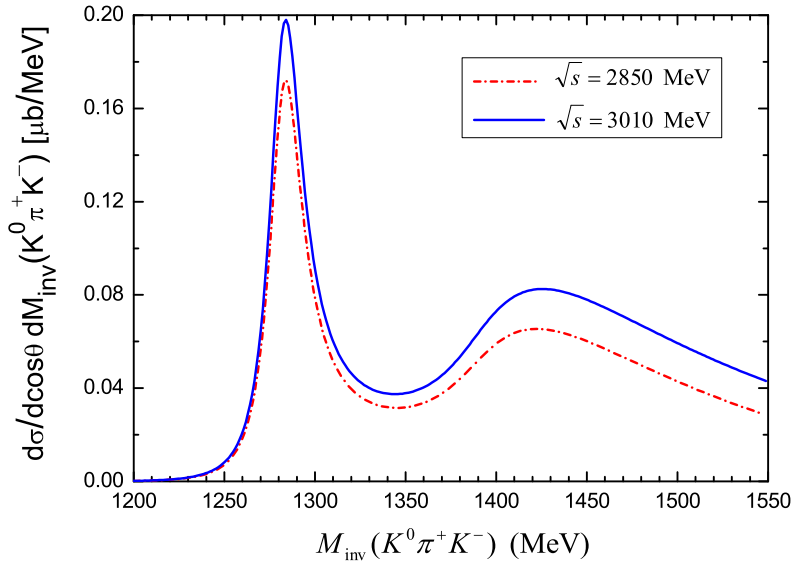}
\end{center}
\vspace{-0.7cm}
\caption{Differential cross section for the $K^- p \to \Lambda K^0 \pi^+ K^-$ reaction for $\cos \!\theta =1$ and $\sqrt{s}=2850$ MeV, $3010$ MeV.}
\label{Fig:4}
\end{figure}

In Fig.~\ref{Fig:4} we show the cross section for the $K^- p \to \Lambda K^0 \pi^+ K^-$ reaction for $\cos \!\theta =1$ and $\sqrt{s}=2850$ MeV, $3010$ MeV.
The results are very similar to those shown before except that they are about a factor of $6$ smaller in size,
according to Eq.~\eqref{eq:cB}.
Once again, the peak at $1420$ MeV appears as before and the ratio of the strengths at the peaks is also very similar to that found before.
We have also evaluated the results with the $BBV$ input of Eqs.~\eqref{eq:vertex}, \eqref{eq:cB} and the form factor normalized to $1$ at $q^2=0$ and, as before,
we find the same shape for the cross section and a size about $\frac{1}{2}$ the former one.

The size of the cross sections obtained is relatively large,
and the fact that there are results in Ref.~\cite{Armenteros} guarantees its measurability in present facilities.

It is worth making some observation about possible contamination of background from other sources.
In order to minimize this potential contribution,
one should bear in mind that the characteristics of the peak at $1420$ MeV that we obtain is that it corresponds to $K^* \bar K$,
concretely $K^{*+} K^-$ and $\bar K^{*0} K^0$ since, as discussed above,
the peak comes from placing the $K^*$ on shell in the diagrams of Figs.~\ref{Fig:1}(b) and \ref{Fig:1}(g).
The other important feature is that the $K^*\bar K$ are produced in relative $s$-wave.
These two conditions can serve as a fitter of other possible background contributions that could distort the picture that we have obtained.

\section{Conclusions}
We have evaluated the cross section for the $\bar K p\to Y K^0 \pi^+ K^-$ reactions,
with $\bar K= \bar K^0, K^-$ and $Y=\Lambda, \Sigma^0, \Sigma^+$ in the region of invariant masses around $\minv (K^0 \pi^+ K^-) \in [1200, 1550]$ MeV.
The fact that the $f_1(1285)$ couples very strongly to the $K^* \bar K$ components
has as a consequence that the mechanism with $K^*$ exchange leads to the production of that resonance in these reactions,
which has been already observed in the $K^- p \to \Lambda f_1(1285)$ reaction.
The novelty presented here is that together with the $f_1(1285)$ excitation, the reaction shows a peak at $1420$ MeV in all these reactions,
which would be very interesting to observe experimentally.
The study done here indicates that this peak, seen in the $K^* \bar K$ final state,
which has been associated to the ``$f_1(1420)$" resonance in different reactions,
is not a resonance but the manifestation of the $K^* \bar K$ decay mode of the $f_1(1285)$.
The observation of these predictions in future experiments would help greatly in clarifying this issue.

\section{ACKNOWLEDGEMENT}
This work is partly supported by the National Natural Science Foundation of China under Grants No. 11975083, No. 11947413,
No. 11847317 and No. 11565007. This work is also partly supported by the Spanish Ministerio de
Economia y Competitividad and European FEDER funds under Contracts No. FIS2017-
84038-C2-1-P B and No. FIS2017-84038-C2-2-P B, and the Generalitat Valenciana in the
program Prometeo II-2014/068, and the project Severo Ochoa of IFIC, SEV-2014-0398.

%\clearpage


\begin{thebibliography}{}

%\cite{XieLiang1}
\bibitem{XieLiang1}
 T.~Sato, T.~Takahashi, and K.~Yoshimura, Lect.\ Notes Phys.\ {\bf 781}, 193 (2009).

%\cite{XieLiang2}
\bibitem{XieLiang2}
  S.~Kumano,
  %``Spin Physics at J-PARC,''
  Int.\ J.\ Mod.\ Phys.\ Conf.\ Ser.\  {\bf 40}, 1660009 (2016)
%  doi:10.1142/S2010194516600090
%  [arXiv:1504.05264 [hep-ph]].

%\cite{XieLiang3}
\bibitem{XieLiang3}
 S.~Okada {\it et al.},
  %``Precision spectroscopy of kaonic atoms at DAPHNE,''
  EPJ Web Conf.\  {\bf 3}, 03023 (2010).
%  doi:10.1051/epjconf/20100303023

%\cite{XieLiang4}
\bibitem{XieLiang4}
 M.~Bazzi {\it et al.} [SIDDHARTA Collaboration],
  %``Preliminary study of kaonic deuterium X-rays by the SIDDHARTA experiment at DAFNE,''
  Nucl.\ Phys.\ A {\bf 907}, 69 (2013).
%  doi:10.1016/j.nuclphysa.2013.03.001
%  [arXiv:1302.2797 [nucl-ex]].

%\cite{Compass}
\bibitem{Compass}
P.~Abbon {\it et al.} [COMPASS Collaboration],
  %``The COMPASS experiment at CERN,''
  Nucl.\ Instrum.\ Meth.\ A {\bf 577}, 455 (2007).
%  doi:10.1016/j.nima.2007.03.026
%  [hep-ex/0703049].

%\cite{Comexo}
\bibitem{Comexo}
 M.~Alekseev {\it et al.} [COMPASS Collaboration],
  %``Observation of a J**PC = 1-+ exotic resonance in diffractive dissociation of 190-GeV/c pi- into pi- pi- pi+,''
  Phys.\ Rev.\ Lett.\  {\bf 104}, 241803 (2010).
%  doi:10.1103/PhysRevLett.104.241803
%  [arXiv:0910.5842 [hep-ex]].

%\cite{Wallner}
\bibitem{Wallner}
  S.~Wallner [COMPASS Collaboration],
  %``Strange-Meson Spectroscopy at COMPASS,''
  arXiv:1911.13079 [hep-ex].

%\cite{Jeff}
\bibitem{Jeff}
 W.~J.~Briscoe, M.~D{\"o}ring, H.~Haberzettl, D.~M.~Manley, M.~Naruki, I.~I.~Strakovsky and E.~S.~Swanson,
  %``Physics opportunities with meson beams,''
  Eur.\ Phys.\ J.\ A {\bf 51}, no. 10, 129 (2015).
%  doi:10.1140/epja/i2015-15129-5
%  [arXiv:1503.07763 [hep-ph]].

%\cite{proposal}
\bibitem{proposal}
Proposal for Jlab PAC47, ``Strange Hadron Spectroscopy with Secondary $K_L$ Beam in Hall D'', S. Adhikari {\it et al.}

%\cite{XieLiang}
\bibitem{XieLiang}
 J.~J.~Xie, W.~H.~Liang and E.~Oset,
  %``$\bar K$-induced formation of the $f_0(980)$ and $a_0(980)$ resonances on proton targets,''
  Phys.\ Rev.\ C {\bf 93}, no. 3, 035206 (2016).
%  doi:10.1103/PhysRevC.93.035206
%  [arXiv:1512.01888 [nucl-th]].

%\cite{pdg}
\bibitem{pdg}
  M.~Tanabashi {\it et al.} [Particle Data Group],
  %``Review of Particle Physics,''
  Phys.\ Rev.\ D {\bf 98}, no. 3, 030001 (2018).
%  doi:10.1103/PhysRevD.98.030001

%\cite{Serpu}
\bibitem{Serpu}
 S.~I.~Bityukov {\it et al.},
  %``Investigation of $D(1285)$ and $e(1420)$ Mesons Production in Exclusive Interactions of $\pi^-$ and $K^-$ Mesons at 32.5-{GeV}/$c$,''
  Sov.\ J.\ Nucl.\ Phys.\  {\bf 39}, 735 (1984)
  [Yad.\ Fiz.\  {\bf 39}, 1165 (1984)].

%\cite{ViniFran}
\bibitem{ViniFran}
  V.~R.~Debastiani, F.~Aceti, W.~H.~Liang and E.~Oset,
  %``Revising the $f_1(1420)$ resonance,''
  Phys.\ Rev.\ D {\bf 95}, no. 3, 034015 (2017).
%  doi:10.1103/PhysRevD.95.034015
%  [arXiv:1611.05383 [hep-ph]].

%\cite{Lutz}
\bibitem{Lutz}
M.~F.~M.~Lutz and E.~E.~Kolomeitsev,
  %``On meson resonances and chiral symmetry,''
  Nucl.\ Phys.\ A {\bf 730}, 392 (2004).
%  doi:10.1016/j.nuclphysa.2003.11.009
%  [nucl-th/0307039].

%\cite{Luis}
\bibitem{Luis}
  L.~Roca, E.~Oset and J.~Singh,
  %``Low lying axial-vector mesons as dynamically generated resonances,''
  Phys.\ Rev.\ D {\bf 72}, 014002 (2005).
%  doi:10.1103/PhysRevD.72.014002
%  [hep-ph/0503273].

%\cite{Geng}
\bibitem{Geng}
Y.~Zhou, X.~L.~Ren, H.~X.~Chen and L.~S.~Geng,
  %``Pseudoscalar meson and vector meson interactions and dynamically generated axial-vector mesons,''
  Phys.\ Rev.\ D {\bf 90}, no. 1, 014020 (2014).
%  doi:10.1103/PhysRevD.90.014020
%  [arXiv:1404.6847 [nucl-th]].

%\cite{Juan}
\bibitem{Juan}
C.~Garcia-Recio, L.~S.~Geng, J.~Nieves and L.~L.~Salcedo,
  %``Low-lying even parity meson resonances and spin-flavor symmetry,''
  Phys.\ Rev.\ D {\bf 83}, 016007 (2011).
%  doi:10.1103/PhysRevD.83.016007
%  [arXiv:1005.0956 [hep-ph]].

%\cite{Aceti}
\bibitem{Aceti}
 F.~Aceti, J.~J.~Xie and E.~Oset,
  %``The $K \bar K \pi$ decay of the $f_1(1285)$ and its nature as a $K^* \bar K -cc$ molecule,''
  Phys.\ Lett.\ B {\bf 750}, 609 (2015).
%  doi:10.1016/j.physletb.2015.09.068
%  [arXiv:1505.06134 [hep-ph]].

%\cite{Birse}
\bibitem{Birse}
M.~C.~Birse,
  %``Effective chiral Lagrangians for spin 1 mesons,''
  Z.\ Phys.\ A {\bf 355}, 231 (1996).
%  doi:10.1007/s002180050105
%  [hep-ph/9603251].

%\cite{Xie}
\bibitem{Xie}
 J.~J.~Xie,
  %``The $K^- p \to f_1(1285) \Lambda$ reaction within an effective Lagrangian approach,''
  Phys.\ Rev.\ C {\bf 92}, no. 6, 065203 (2015).
%  doi:10.1103/PhysRevC.92.065203
%  [arXiv:1509.06196 [nucl-th]].

%\cite{Kaiser}
\bibitem{Kaiser}
F.~Klingl, N.~Kaiser and W.~Weise,
  %``Current correlation functions, QCD sum rules and vector mesons in baryonic matter,''
  Nucl.\ Phys.\ A {\bf 624}, 527 (1997).
%  doi:10.1016/S0375-9474(97)88960-9
%  [hep-ph/9704398].

%\cite{RamosVec}
\bibitem{RamosVec}
  E.~Oset and A.~Ramos,
  %``Dynamically generated resonances from the vector octet-baryon octet interaction,''
  Eur.\ Phys.\ J.\ A {\bf 44}, 445 (2010).
%  doi:10.1140/epja/i2010-10957-3
%  [arXiv:0905.0973 [hep-ph]].

%\cite{SakaiRamos}
\bibitem{SakaiRamos}
  S.~Sakai, E.~Oset and A.~Ramos,
  %``Triangle singularities in $B^-\rightarrow K^-\pi^-D_{s0}^+$ and $B^-\rightarrow K^-\pi^-D_{s1}^+$,''
  Eur.\ Phys.\ J.\ A {\bf 54}, no. 1, 10 (2018).
%  doi:10.1140/epja/i2018-12450-5
%  [arXiv:1705.03694 [hep-ph]].

%\cite{Doring}
\bibitem{Doring}
  M.~Doring, C.~Hanhart, F.~Huang, S.~Krewald, U.-G.~Meissner and D.~Ronchen,
  %``The reaction pi+ p --> K+ Sigma+ in a unitary coupled-channels model,''
  Nucl.\ Phys.\ A {\bf 851}, 58 (2011).
%  doi:10.1016/j.nuclphysa.2010.12.010
%  [arXiv:1009.3781 [nucl-th]].

%\cite{Ronchen}
\bibitem{Ronchen}
  D.~Ronchen {\it et al.},
  %``Coupled-channel dynamics in the reactions piN --> piN, etaN, KLambda, KSigma,''
  Eur.\ Phys.\ J.\ A {\bf 49}, 44 (2013).
%  doi:10.1140/epja/i2013-13044-5
%  [arXiv:1211.6998 [nucl-th]].

%\cite{Holinde}
\bibitem{Holinde}
  R.~Machleidt, K.~Holinde and C.~Elster,
  %``The Bonn Meson Exchange Model for the Nucleon Nucleon Interaction,''
  Phys.\ Rept.\  {\bf 149}, 1 (1987).
%  doi:10.1016/S0370-1573(87)80002-9

%\cite{XieZou}
\bibitem{XieZou}
  J.~J.~Xie and B.~S.~Zou,
  %``The Role of Delta++*(1620) resonances in pp ---> nK+ Sigma+ reaction and its important implications,''
  Phys.\ Lett.\ B {\bf 649}, 405 (2007).
%  doi:10.1016/j.physletb.2007.04.035
%  [nucl-th/0701021].

%\cite{Armenteros}
\bibitem{Armenteros}
 A.~Gurtu {\it et al.} [Amsterdam-CERN-Nijmegen-Oxford Collaboration],
  %``Production and Decay Properties of the $D(1285)$ Meson in $K^- p$ Interactions at 4.2-{GeV}/$c$,''
  Nucl.\ Phys.\ B {\bf 151}, 181 (1979).
%  doi:10.1016/0550-3213(79)90433-4


\end{thebibliography}
\end{document}